# Exploring Thixotropic Timescale: Phenomenological Insights and Analytical Perspectives

Yogesh M. Joshi[1, 2, 3, *]

[1] Department of Chemical Engineering, Indian Institute of Technology Kanpur, Kanpur, Uttar Pradesh 208016, India.

[2] Materials Science Programme, Indian Institute of Technology Kanpur, Kanpur, Uttar Pradesh 208016, India.

[3] Centre for Nanosciences, Indian Institute of Technology Kanpur, Kanpur, Uttar Pradesh 208016, India.

* Email: joshi@iitk.ac.in

**Abstract**

Thixotropy is characterized by an increase in viscosity when a material is subjected to no flow (quiescent) or weak flow conditions and a decrease in viscosity when it is subjected to strong flow conditions. The characteristic timescale associated with the thixotropic phenomenon, particularly how the viscosity increases with time, has been termed the thixotropic timescale. In the literature, several approaches have been suggested for estimating the thixotropic timescale. The most prominent approach, however, infers it from a specific form of a kinetic expression for structure parameter evolution. In this paper, we study the various kinds of structural kinetic models, and by carrying out a careful analysis of the same, we propose a parameter for the thixotropic timescale that is associated with the most generic form of the kinetic expression for structure parameter evolution. We observe that when the viscosity of the structural kinetic model undergoes continuous increase with time and eventually diverges under quiescent conditions, which we believe is the most practical scenario, our analysis suggests that increasing the thixotropic timescale weakens the thixotropic character of a system. We also propose a new phenomenological measure of the thixotropic timescale: $\tau_{thix} = (d\ln\eta/dt)^{-1}$, where $\eta$ is viscosity and $t$ is time. The proposed definition allows a straightforward and unique way to determine thixotropic timescale through experiments and agrees well with the conventional notion of thixotropy.

**Introduction**

Thixotropy is associated with an increase in the viscosity of a material when it is subjected to no flow (quiescent) or weak flow conditions and a decrease in viscosity under strong flow conditions [1]. Distinctly, the phenomenon is related to two aspects: one is an increase in viscosity, which is an outcome of microstructural build-up; while the second is a decrease in viscosity that occurs due to microstructural breakdown under flow [2-4]. The characteristic time associated with the structural build-up that causes an increase in viscosity may typically be termed the thixotropic or restructuration timescale [5-7]. The other timescale is an imposed timescale associated with the applied deformation field, typically an inverse of the second invariant of the rate of strain tensor [1]. The ratio of these two timescales is a natural dimensionless number, the value of which is expected to determine how the microstructure gets altered as a function of time under the applied deformation field [7-10]. In this work, we discuss various features associated with the restructuration or thixotropic timescale, and possible issues with the way this timescale is presently defined. Our analysis leads to a proposal for a thixotropic timescale obtained from a generic kinetic expression of the evolution of structure parameter. We also put forth a phenomenological proposition of the thixotropic timescale, offering a straightforward experimental estimation method that agrees well with the established understanding of thixotropy.

According to Barnes [11], the origin of the term thixotropy is attributed to Peterfi [12]. The historic development associated with the same has been discussed in detail elsewhere [7, 11]. According to IUPAC, thixotropy is defined as [7]: *"the continuous decrease of viscosity with time when the flow is applied to a sample that has been previously at rest, and the subsequent recovery of viscosity when the flow is discontinued."* As comprehensively articulated in the literature [3, 7, 10, 11, 13, 14], this definition does not safeguard the inclusion of non-linear viscoelasticity under the purview of thixotropy. Consequently, doubts have been raised about the independent existence of the phenomenon of thixotropy, and if it does have an independent existence, boundaries have been created to unequivocally separate it from nonlinear viscoelasticity [7, 13]. In our opinion, any material capable of showing microstructural evolution that causes increase in viscosity under no-flow and no stress conditions, and is homogeneous over mesoscopic lengthscales, such that the whole phenomenon is deformation field reversible is intrinsically thixotropic in nature. However, distinction between thixotropy and viscoelasticity, and what issues it leads to, is discussed elsewhere [3, 7, 10, 11, 13-15], and hence we will not dwell on it in this study.

In the contemporary literature, there is no unique way to explicitly describe the thixotropic timescale. Some definitions employ the structural kinetic model to illustrate the thixotropic timescale, while others infer it from the experimental data, either by associating it with a specific experimental feature or by simply fitting the experimental data with generic expressions. The structural kinetic model has been considered to be the most convenient representation to introduce the thixotropic timescale [5]. It is comprised of an evolution equation of structure parameter that denotes a conceptual dimensionless measure of a structure at a given time instance. However, since there are no strict measures that put constraints on the variation of structure parameter, one class of models constrains the structure parameter between 0 and 1 [13], while the other class of models allows variation of the same between 0 and $\infty$ [6, 16]. In both cases, the lower limit is the thixotropic structureless state, while the upper limit is a full or equilibrium structured state. If the structure parameter in the former case is assumed to be $\lambda \in (0, 1)$, the evolution of the same in the absence of any flow (quiescent conditions) typically takes a form: $d\lambda/dt = (1 - \lambda)/T_1$, where $t$ is time and $T_1$ is a constant having dimensions of time [5]. Therefore, for this formalism, with an initial state as a structureless state, $\lambda$ evolves as: $\lambda = 1 - \exp(-t/T_1)$. In this expression $\lambda$ increases with increase in $t/T_1$, which is suggestive of structural build-up. Consequently, the corresponding timescale $T_1$ has been termed as the thixotropic time scale [5, 7, 8, 10]. Interestingly, the literature is mute what $T_1$ should be called if one considers a different form $d\lambda/dt = \mathcal{F}(\lambda)/T_1$ under quiescent conditions, where $\mathcal{F}(\lambda)$ is an arbitrary function of $\lambda$, for example $\mathcal{F}(\lambda) = (1 - \lambda)^p$, $p > 0$.

We now consider a case wherein the structure parameter is unbounded: $\Lambda \in (0, \infty)$. For this case, the evolution equation of the same in the absence of any flow has been proposed to have a form $d\Lambda/dt = 1/T_0$ [6]. In this case, $T_0$ also has dimensions of time. Staring from the structureless state, $\Lambda$ evolves as $\Lambda = t/T_0$. In this expression, $\Lambda$ increases as $t/T_0$ increases, indicating structural growth. Consequently, the corresponding timescale $T_0$ has been termed as the characteristic time of structural evolution or the restructuration time [6, 7], which has identical physical meaning as the thixotropic timescale. Interestingly, depending on whether a material reaches finite viscosity associated with an equilibrium state or infinite viscosity by undergoing eternal aging, two seemingly different measures of thixotropic timescales have been proposed in the literature [7]. Furthermore, as discussed in greater detail below, depending on the bounds of the structure parameter, whether an increase or decrease in the so-called thixotropic time causes weakening or strengthening of the thixotropic character of a material, is also a matter of contention. There are several issues that need careful assessment. The first one is, can a structure parameter, which by itself does not have any definite physical meaning, be used to describe the thixotropic timescale? Secondly, can the definition of thixotropic timescale, which in principle, could be an important characteristic feature of a thixotropic material, be confined to a specific form of evolution equation? Another issue is, whether thixotropic time is a constant or can vary with time. In this work, we shall analyze these questions in detail.

There have been various other proposals wherein the thixotropic timescale has been inferred from the experimental data. Ewoldt and coworkers [9] proposed a procedure that involves the application of step change in shear rate or magnitude of oscillatory strain to a thixotropic material and recording the corresponding output variable (either stress or elastic modulus). Subsequently, they fitted an appropriate function involving the Maxwell model like a sum of exponential decay, or the Kelvin-Voigt model like a sum of retarded increase to the output variable, whichever is applicable. The corresponding fitted timescale leads to either a discrete or continuous spectrum (by converting the sum to an integral) of characteristic timescales involved in the process. They term these spectra to be recovery or breakdown spectra depending upon the whether process of applied step jump involves recovery or breakdown in the structure. In another proposal, the thixotropic timescale has also been inferred from the rheological hysteresis experiments. In a typically adapted procedure, thixotropic material is subjected to a stepwise decrease in the shear rate from a high value to a low value followed by a stepwise increase in the shear rate back to the high value with a certain step time $\delta t$. It has been observed that the area of the hysteresis loop when plotted as a function of $\delta t$ shows a bell-shaped curve. The characteristic thixotropic timescale of a material is proposed to be related to the value of $\delta t$ at which the area of the hysteresis loop is maximum [8, 17-20]. Interestingly, hysteresis loops are also observed for non-linear viscoelastic materials and hence they do show the bell-shaped curve when the area of the hysteresis loop is plotted as a function of $\delta t$ [21]. Consequently, there is always a concern regarding how much the viscoelastic character of a material influences the nature of the bell-shaped curve in thixo-viscoelastic materials.

With respect to the aforementioned background, we analyze the following aspects in this work. Firstly, we assess a question of whether the definition of thixotropic timescale should depend on how a structural kinetic model has been proposed. We consider two different kinds of structural kinetic models and discuss various features of the thixotropic timescale they lead to. We discuss an important aspect of limits on the thixotropic timescale and its relation to the weakening/strengthening of the thixotropic phenomenon. Subsequently, we consider a most generic form of the evolution equation of the structure parameter and propose a specific parameter as a universal candidate for the thixotropic time scale. Finally, taking a cue from this discussion and considering the intrinsic definition of thixotropy, we also propose a new phenomenological measure of thixotropic timescale and discuss characteristics associated with the same.

**Thixotropic timescales from the structural kinetic models**

The structural kinetic model is generally thought to be the most useful theoretical framework for describing thixotropic behavior. The structural kinetic model is comprised of three components [22-24]. The first component is the evolution of the structure parameter that conceptually describes the actual state of the structure. Usually, the evolution equation of the structure parameter is represented by first-order kinetics with microstructural build-up and break-down terms. The second component is the constitutive equation, which could be inelastic or viscoelastic with or without frame invariance depending upon the application. Finally, the third and equally important component is the relationship between the structure parameter and the parameters of the constitutive equation such as viscosity, modulus, etc. The third component is the key that translates the conceptual but still vague notion of a state of the structure to more tangible and measurable properties of a material. One of the earliest structural kinetic models is due to Goodeve and Whitfield [25], whose more generalized form can be expressed as:

$$\frac{d\lambda}{dt} = \frac{(1-\lambda)}{T_1} - f(\lambda)\dot{\gamma}. \tag{1}$$

The first term on the right denotes the microstructural build-up while the second term depicts the microstructural break-down. The specific form proposed by Goodeve and Whitfield [25] is obtained for $f(\lambda) = k\lambda$. For the class of expression given by Eq. (1), $\lambda$ varies from 0 (devoid of structure) to 1 (completely structured or equilibrium state). The microstructural breakdown is due to the applied deformation field whose strength is expressed by the second invariant of the rate of strain tensor given by $\dot{\gamma}$. Dimensionally, the parameter $T_1$ has units of inverse of time, and since it signifies the rate of structural evolution, $T_1$ has been termed the thixotropic timescale in the literature [5, 7, 8, 10]. If a constant shear rate $\dot{\gamma}_0$ is applied to a material following Eq. (1) with $f(\lambda) = k\lambda$, in its structureless state ($\lambda = 0$ at $t = 0$), Eq. (1) can be solved analytically to obtain,

$$\lambda = \frac{1 - exp(-(kT_1\dot{\gamma} + 1)(t/T_1))}{1 + kT_1\dot{\gamma}} \qquad (2)$$

It can be seen that Eq. (2) intrinsically leads to a dimensionless group: $T_1\dot{\gamma}$, which has been termed as Thixotropy number by Mujumdar et al. [5] and Larson and Wei [7], while the Mnemosyne number by Jamali and McKinley [8]. Interestingly Eq. (2) also leads to another dimensionless number, $t/T_1$. Since $T_1$ has been expressed as the thixotropic timescale, the dimensionless group $t/T_1$ has been represented as mutation number [8] or an inverse of the thixotropic counterpart of the Deborah number [10].

In Fig. 1(a), we plot $\lambda$ as a function of $t/T_1$ for different values of $T_1\dot{\gamma}$ for $k = 1$. In a limit of $t/T_1 \to 0$, $\lambda$ increases with $t/T_1$. For $T_1\dot{\gamma} \gg 1$, $\lambda$ reaches a steady state for $t/T_1 \approx (T_1\dot{\gamma})^{-1}$, while for $T_1\dot{\gamma} \ll 1$, $\lambda$ attains a steady state for $t/T_1 \approx 1$. Furthermore, as evident from Eq. (2), for $T_1\dot{\gamma} \gg 1$ steady state value of $\lambda$ is given by $\lambda_{ss} \approx (T_1\dot{\gamma})^{-1}$. On the other hand, for $T_1\dot{\gamma} \ll 1$, $\lambda_{ss}$ saturates to 1 with decrease in $T_1\dot{\gamma}$. Furthermore, in this representation, it is proposed that in the limit $T_1\dot{\gamma} \ll 1$ effect of thixotropy can be neglected while in the limit $T_1\dot{\gamma} \gg 1$, the effect of thixotropy is strong [5, 8, 10, 26]. On the other hand, on similar lines, it has been suggested that the limit of $T_1 \to 0$ is devoid of thixotropy. Under quiescent conditions ($\dot{\gamma} = 0$), since it takes $t = \mathcal{O}(T_1)$ time for $\lambda$ to reach the equilibrium, in a limit of $T_1 \to 0$, material reaches the full structured state (the equilibrium state) from the structureless state instantaneously. Thereafter material remains in the time-invariant state forever. Consequently, in terms of the so-called thixotropic Deborah number, any finite time larger than $T_1$ ($t \gg T_1$) is suggestive of an extremely weak (or no) thixotropy limit in terms of time. For a thixo-viscoelastic material, ratio of stress relaxation time to thixotropic time has been termed as the thixoelastic parameter [8, 26]. It has been proposed that when the thixotropic timescale is far greater than the stress relaxation time, thixotropic effects dominate over the viscoelastic effects and the other way around. In general, it is clear that in this representation, under quiescent conditions, the thixotropic timescale is perceived as similar to that of viscoelastic relaxation timescale but in the thixotropic context. The way in the stress relaxation experiments, a viscoelastic material achieves the equilibrium after time of the order of relaxation time has passed, thixotropic material is expected to attain the equilibrium for times greater than the thixotropic timescale.

Let us now study an alternative representation. In an important contribution, Coussot et al. [6, 16] proposed a different structural kinetic model given by:

$$\frac{d\Lambda}{dt} = \frac{1}{T_0} - F(\Lambda)\dot{\gamma}, \qquad (3)$$

where $\Lambda$ is the newly defined structure parameter that varies from $0$ (devoid of structure) to $\infty$ (progressively higher value of $\Lambda$ suggests build-up of structure to a greater extent). Interestingly, for eternally aging materials, Larson and Wei [7] proposed a similar kinetic equation for a structure parameter ($\Lambda_1$) given by:

$$\frac{d\Lambda_1}{dt} = \frac{1}{T_0} - \Lambda_1 F_1(\dot{\gamma}), \qquad (4)$$

In both Eq. (3) and (4), $T_0$ has dimensions of a timescale. Coussot defines $T_0$ as the restructuration time of a material [6], while Larson and Wei [7] term the product $T_0 F_1(\dot{\gamma})$ as thixotropy number, implying $T_0$ to be the thixotropic timescale. A specific case of Eq. (3) has been proposed by Coussot et al. [6, 16] with $F(\Lambda) = \beta\Lambda$, where $\beta$ is a constant. Interestingly if we consider $F_1(\dot{\gamma}) = \beta\dot{\gamma}$ in Eq. (4), the model by Larson and Wei [7] (with $\Lambda = \Lambda_1$) becomes identical to that proposed by Coussot et al. (Eq. (3)) [6, 16]. If we subject the material following Eq. (3) in its structureless state ($\lambda = 0$ at $t = 0$) to a constant shear rate $\dot{\gamma}_0$, we get,

$$\Lambda = \frac{1 - exp\bigl(-\beta(T_0\dot{\gamma}_0)(t/T_0)\bigr)}{\beta T_0 \dot{\gamma}}. \qquad (5)$$

As per Eq. (5), as $T_0 \to 0$, $\Lambda$ goes on increasing at a faster rate. In this approach also we get two dimensionless numbers: $T_0\dot{\gamma}$ and $T_0/t$. Out of which the former is equally good candidate for thixotropy number.

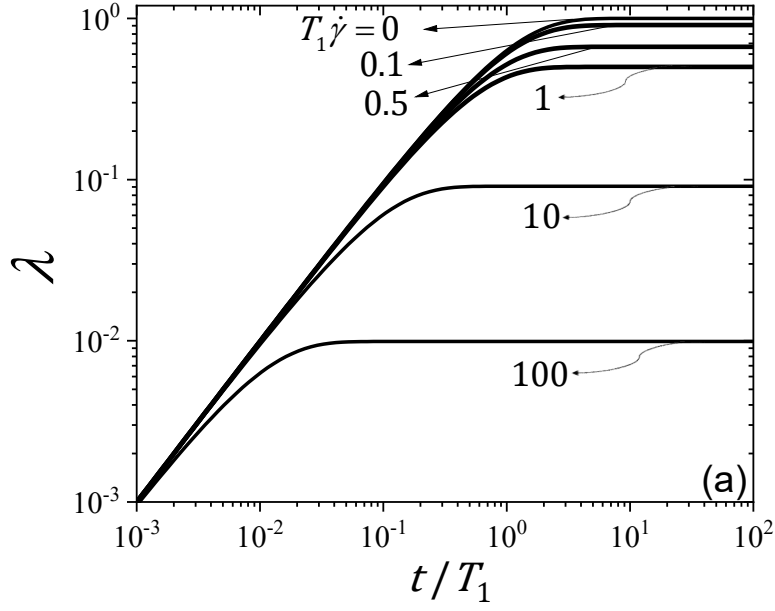

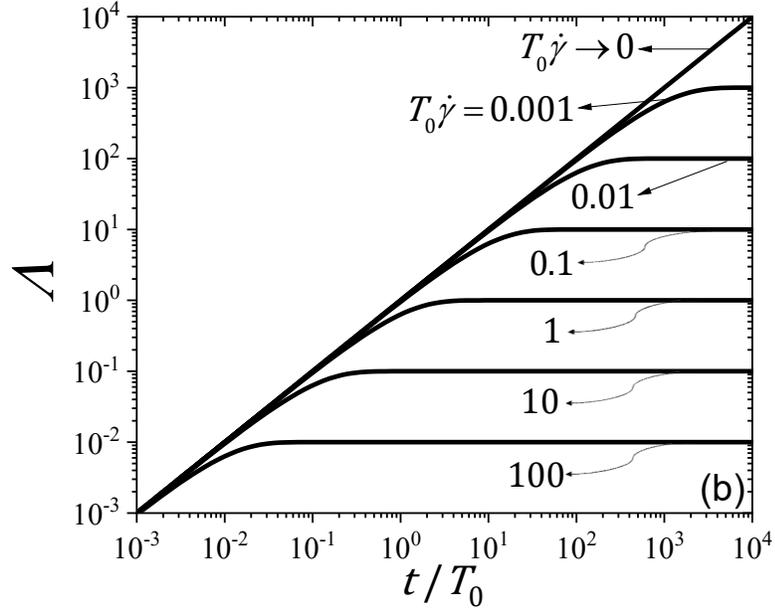

**Figure 1.** Structure parameter is plotted as a function of dimensionless time for different values of dimensionless shear rate ($T_1 \dot{\gamma}$ or $T_0 \dot{\gamma}$). (a) represents Eq. (2) with $k = 1$, wherein $\lambda$ is plotted as a function of $t/T_1$ while (b) represents Eq. (5) with $\beta = 1$, wherein $\Lambda$ is plotted as a function of $t/T_0$.

In Fig. 1(b) we plot $\Lambda$ as a function of $t/T_0$ for different values of $T_0 \dot{\gamma}$ for $\beta = 1$. In a limit of $t/T_0 \to 0$, $\Lambda$ increases with $t/T_0$. Interestingly, in this formalism, regardless of any non-zero value of $T_0 \dot{\gamma}$, $\Lambda$ reaches a steady state for $t/T_0 \approx (T_0 \dot{\gamma})^{-1}$ and the steady state value of $\Lambda$ is given by $\Lambda_{ss} \approx (T_0 \dot{\gamma})^{-1}$. Consequently, for any given value of $\dot{\gamma}$, the value of $\Lambda$ will take a higher value with a decrease in $T_0$ when compared at the same $t$. Most distinctly, for $T_0 \dot{\gamma} \to 0$, $\Lambda$ keeps on increasing indefinitely and follows: $\Lambda \sim t/T_0$, which suggests at any value of $t$, $\Lambda$ increases with a decrease in $T_0$ to take progressively higher values and tends to $\infty$. In any case, a decrease in $T_0$ causes an increase in $\Lambda$, which, by definition, means a greater extent of microstructural build-up. Consequently, under quiescent conditions, in this formalism $T_0 \to 0$ is not a progressive weakening of thixotropic behavior. On the contrary, according to Eq. (3) and (4), as $T_0$ decreases, $d\Lambda/dt$ goes on increasing. Consequently, rate of structure formation becomes faster. Thixotropic effect then can actually be considered to be getting stronger in a limit of $T_0 \to 0$. On the other hand, in Eq. (3), in a limit $T_0 \to \infty$, $d\Lambda/dt \to 0$ and the rate of enhancement of $\Lambda$ or the rate of structural build-up gets weaker. Hence, in a limit of $T_0 \to \infty$ effect of thixotropy can be considered to be weakening. As per Eq. (5), for $T_0 \to \infty$ is a limit for which $\Lambda$ remains at its structureless value forever ($\Lambda = 0$), suggesting material is devoid of thixotropy. In this framework, $T_0$ does not come across as a thixotropic counterpart of relaxation time. Consequently, the ratio $T_0/t$ may not be termed as thixotropic counterpart of Deborah number. The only aspect one can be certain about is higher the value of $T_0/t$ weaker is the microstructural evolution.

The dichotomy associated with the thixotropic timescale defined in two different ways appears to originate because of different bounds on the structure parameter, which as such is a mere representative description of a state of the structure. Consequently, what truly matters is the relationship between the parameters of the constitutive equation (such as viscosity, elastic modulus, etc.) and the structure parameter. Furthermore, we need to obtain a relationship between above defined two kinds of thixotropic timescales for the identical rheological behavior shown by the corresponding models denoted by Eq. (1) and Eq. (3) (or Eq. (4)). Firstly, we assume that the rheological response is inelastic and is described by the generalized Newtonian model given by:

$$\boldsymbol{\sigma} = \eta \dot{\boldsymbol{\gamma}}, \tag{6}$$

where $\dot{\gamma}$ is the rate of strain tensor and $\sigma$ is the stress tensor. The viscosity ($\eta$) is a function of the structure parameter given by: $\eta = \eta_0 h(\lambda) = \eta_0 H(\Lambda)$, such that for $\lambda = 0$ or $\Lambda = 0$, $\eta = \eta_0$. We now transform Eq. (3) from $\Lambda \in (0, \infty)$ to $\lambda \in (0,1)$ using $\Lambda = -\ln(1-\lambda)$ leading to,

$$\frac{d\lambda}{dt} = \frac{(1-\lambda)}{T_0} - (\lambda - 1)F(-\ln(1-\lambda))\dot{\gamma}. \tag{7}$$

Interestingly Eq. (7) takes the same form as Eq. (1) with $f(\lambda) = (\lambda - 1)F(-\ln(1-\lambda))$ and $T_1 = T_0$. On the other hand, Eq. (1) can be transformed from $\lambda \in (0,1)$ to $\Lambda \in (0, \infty)$ using $\lambda = 1 - e^{-\Lambda}$ to give:

$$\frac{d\Lambda}{dt} = \frac{1}{T_1} - e^{\Lambda} f(1 - e^{-\Lambda})\dot{\gamma}. \tag{8}$$

On similar lines, Eq. (8) takes the same form as Eq. (3) with $F(\Lambda) = e^{\Lambda} f(1 - e^{-\Lambda})$ and $T_0 = T_1$. The thixotropic time scale is related to evolution of material under quiescent conditions ($\dot{\gamma} = 0$). In addition, the thixotropic time scale as defined in the literature [5, 7, 8, 10] for structural kinetic models with $\lambda \in (0,1)$ is associated with the coefficient of the first term on the right-hand side of Eq. (1) or (7) and hence, it is independent of the functional form $f(\lambda)$. On similar lines, thixotropic timescale as defined for structural kinetic models with $\Lambda \in (0, \infty)$ is also independent of the nature of function $F(\Lambda)$ [6, 7]. It is therefore astonishing that the thixotropic timescale comes out to be identical ($T_0 = T_1$) for both the formulations that employ $\lambda \in (0,1)$ to $\Lambda \in (0, \infty)$. It is, therefore, perplexing that two seemingly different formalisms, but with the identical definition of thixotropic timescale, lead to completely opposite limits of thixotropy.

**Two frameworks: Infinite and finite viscosity models**

To explore the matter further we need to analyze how structure parameter is perceived in a limit of the equilibrium state. The case $\lambda \in (0,1)$ seemingly suggests that the fully structured equilibrium state is well defined with $\lambda = 1$. In contrast, for $\Lambda \in (0, \infty)$ the upper limit associated with the fully structured equilibrium state is ill-defined since it gets realized as $\Lambda \to \infty$. However, as previously mentioned, a structure parameter, whether $\lambda \in (0,1)$ or $\Lambda \in (0, \infty)$, signifies a conceptual but still vague notion of the microstructure's state at a given point in time. Consequently, in order to have any tangible information from a structure parameter it must be related to parameters of the constitutive model, which for the present case is viscosity, given by: $\eta = \eta_0 h(\lambda) = \eta_0 H(\Lambda)$. Notwithstanding which structural kinetic model is used, we consider two cases: (i) the equilibrium state has finite viscosity, and the corresponding model is termed a finite viscosity model, and (ii) the equilibrium state has infinite viscosity and the corresponding model is termed as infinite viscosity model.

In order to analyze the effect of viscosity on structure parameter, particularly whether viscosity remains finite ($\eta_{fin}$) or approaches progressively high value tending to infinity ($\eta_{inf}$), we use the following two expressions that have already been proposed in the literature. For $\eta_{fin}$, we use an expression proposed by Jamali and McKinley [8], given by:

$$\eta_{fin} = \eta_0 + \eta_p \lambda = \eta_0 + \eta_p(1 - e^{-\Lambda}), \tag{9}$$

where $\eta_p$ is the structural contribution to the viscosity such that $\eta_0 + \eta_p$ is the equilibrium viscosity associated with a material. In this work we consider $\eta_p/\eta_0 = 100$. We use the same expression for $\eta_{inf}$ as proposed by Coussot et al. [16, 27] and used by Wei et al. [27]:

$$\eta_{inf} = \eta_0 e^{\alpha \Lambda} = \eta_0 e^{-\alpha \ln(1-\lambda)} = \eta_0(1 - \lambda)^{-\alpha}, \tag{10}$$

where $\alpha$ is a positive number. Throughout this work we consider it to be $\alpha = 1$. These models can now be described as, (i) Finite viscosity model: Eq. (3) with with $F(\Lambda) = \beta \Lambda$, Eq. (6), and Eq. (9); and (ii) Infinite viscosity model: Eq. (3) with with $F(\Lambda) = \beta \Lambda$, Eq. (6), and Eq. (10). As discussed above, the choice of structural kinetic parameter evolution equation either Eq. (1) (that is same as Eq. (7)) or Eq. (3) (that is identical to Eq. (8)) is immaterial as through $\lambda \in (0,1) \Leftrightarrow \Lambda \in (0, \infty)$ transformation, we should expect the identical results. It is important to note that, the infinite viscosity model also shows time dependent yield stress through viscosity bifurcation. Other variations of the infinite viscosity model employed elsewhere show monotonic, Herschel–Bulkley-like (constant stress plateau in a limit of small shear rates), or non-monotonic flow curves for the steady-state flow [4, 24, 28, 29]. The expression of viscosity given by Eq. (10) specifically leads to a non-monotonic flow curve, and hence necessarily shows yield stress [16].

To begin with, we solve both the models analytically with an initial condition of $\Lambda = 0$ at $t = 0$, for different values of $T_0 \dot{\gamma}$. The results of the finite viscosity model are shown in Fig. 2(a). It can be seen that the material always attains a steady state under the application of any shear rate (zero or non-zero) that is given by: $\eta_{fin,ss} =$

$\eta_0 + \eta_p\left(1 - e^{-1/\beta T_0 \dot{\gamma}}\right)$. In a limit of very high shear rate ($\dot{\gamma} \to \infty$), the viscosity does not change with time and always remains at the initial value ($\eta_{fin} = \eta_{fin,ss} = \eta_0$). The results can also be interpreted by considering the value of $\dot{\gamma}$ to be constant and varying $T_0$. Accordingly, for a given $\dot{\gamma}$, decrease in $T_0$, on one hand, allows the attainment of a higher value of viscosity but the time required to reach the same goes on decreasing. Increase in $T_0$, on the other hand, reduces the steady state value of viscosity but attainment of the same gets prolonged.

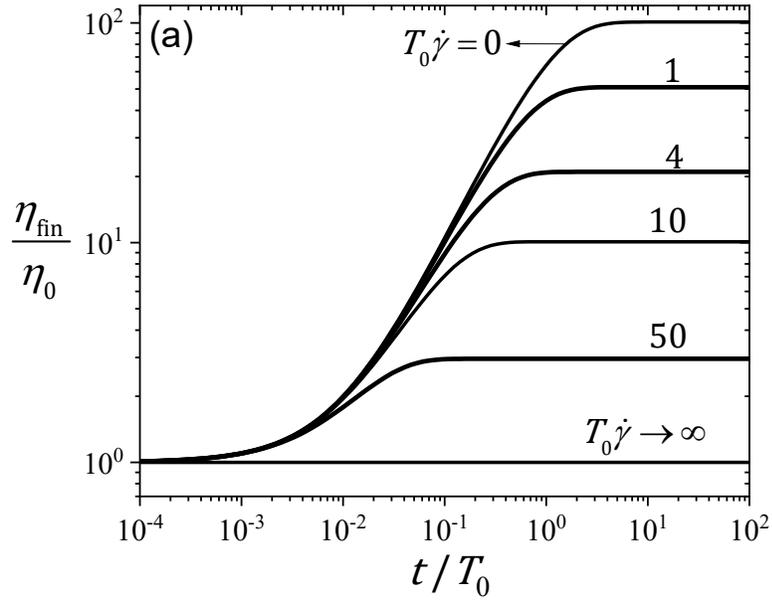

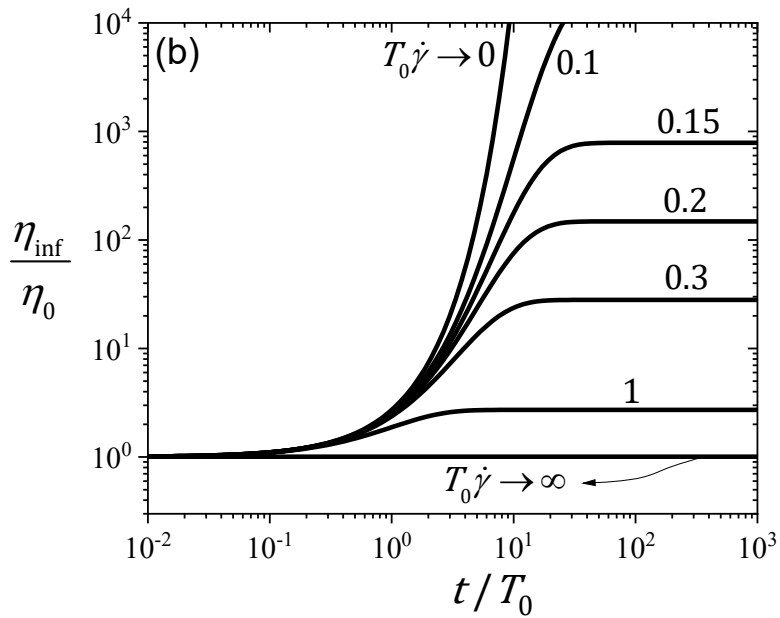

**Figure 2** Viscosity normalized by viscosity associated with the structureless state is plotted for (a) Finite viscosity model ($\eta_p/\eta_0 = 100$) and (b) Infinite viscosity model ($\alpha = 1$) as a function of $t/T_0$ for various values of $T_0\dot{\gamma}$. For both the models we consider $\beta = 1$.

Now we turn to the infinite viscosity model, whose results are plotted in Fig. 2(b). For this model, the material always attains a steady state under the application of any non-zero shear rate that is given by: $\eta_{inf,ss} = \eta_0 e^{\alpha/\beta T_0 \dot{\gamma}}$. As expected, for a very high shear rate ($\dot{\gamma} \to \infty$), the viscosity always remains constant at the initial value ($\eta_{inf} = \eta_{inf,ss} = \eta_0$). Furthermore, for a constant $\dot{\gamma}$, decrease in $T_0$ leads to a higher value of steady state viscosity but the time associated with achievement of the steady state goes on decreasing. Increase in $T_0$ causes a decrease in the steady state value of viscosity, however, it takes a proportionally longer time to attain the same. In a limit of either $T_0 \to 0$ or $\dot{\gamma} \to 0$, the viscosity goes on increasing indefinitely as shown in Fig. 2(b).

Among the various cases discussed in Fig 2, the scenario of quiescent conditions, that is $\dot{\gamma} = 0$ is an important case. Consequently, we solve both models analytically for quiescent or no flow conditions ($\dot{\gamma} = 0$) with an initial condition of $\Lambda = 0$ at $t = 0$. For either of the cases, $\Lambda$ evolves linearly as given by: $\Lambda = t/T_0$. Consequently, for a finite viscosity model, an evolution of viscosity is given by:

$$\frac{\eta_{fin}}{\eta_0} = 1 + \frac{\eta_p}{\eta_0}\left(1 - \exp\left(-\frac{t}{T_0}\right)\right). \tag{11}$$

Eq. (11) clearly shows that in a limit of $t/T_0 \to \infty$, equilibrium value of $\eta_{fin,E} = \eta_0 + \eta_p$ is attained. An evolution of viscosity for an infinite viscosity model is given by:

$$\frac{\eta_{inf}}{\eta_0} = \exp\left(\alpha \frac{t}{T_0}\right). \tag{12}$$

As expected, Eq. (12) suggests that $\eta$ increases indefinitely with increase in time normalized by $T_0$.

In Fig. 3(a) we plot normalized viscosity for the finite viscosity model given by Eq. (11) for different values of thixotropic timescale $T_0$. It can be seen that, in a limit of $T_0 \to 0$ the viscosity instantaneously approaches the equilibrium value and thereafter remains invariant of time. As a result, this limit of $T_0 \to 0$ has been termed as no thixotropy limit. With increase in $T_0$, time required to attain the equilibrium goes on increasing such that equilibrium is achieved over a timescale of $\sigma(T_0)$. Fig. 3(b) depicts the time evolution of normalised viscosity for the infinite viscosity model given by Eq. (12). In a limit of $T_0 \to 0$, the viscosity increases extremely sharply. However, the rate of the time evolution of viscosity reduces as $T_0$ increases, when compared at any time $t$.

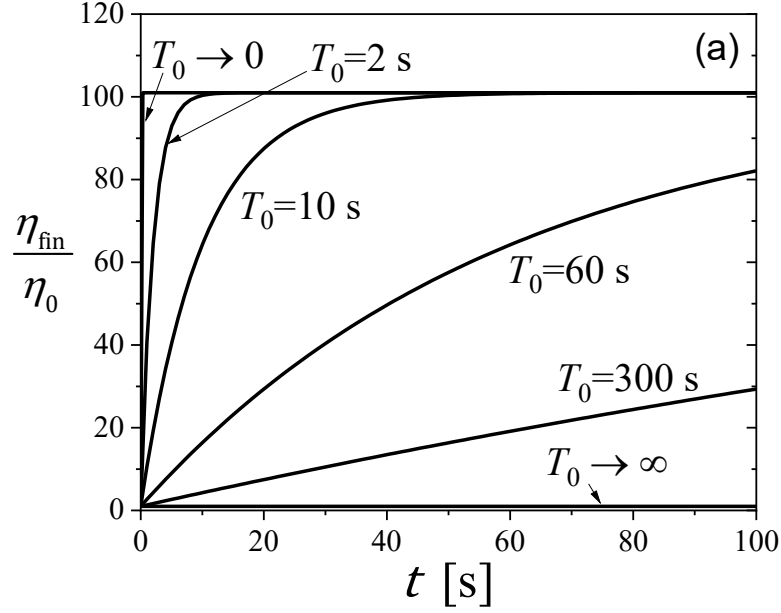

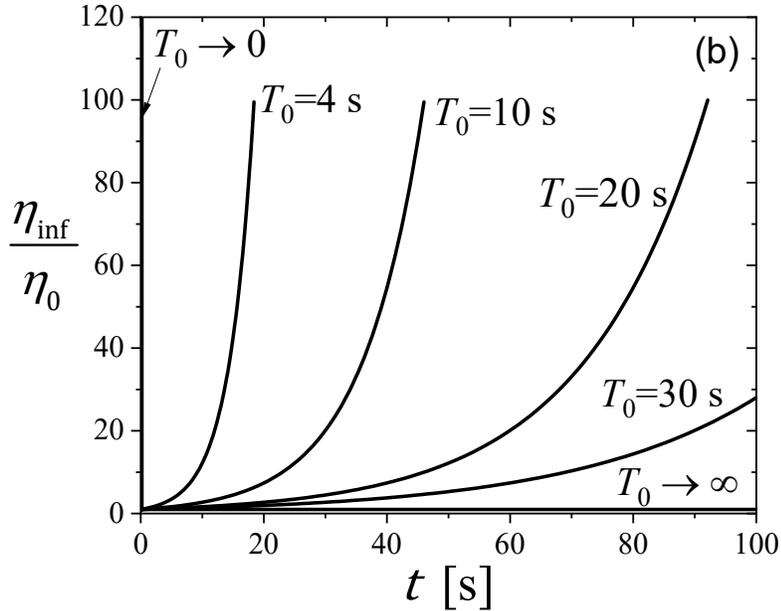

**Figure 3.** (a) Normalized viscosity is plotted as a function of dimensional time for the finite viscosity model given by Eq. (11) for no flow conditions ($\dot{\gamma} = 0$). It can be seen that, with increase in $T_0$, time required to achieve the equilibrium viscosity ($\eta_0 + \eta_p$) goes on increasing. Consequently, thixotropic behavior has been proposed to get stronger with increase in $T_0$. (b) Normalized viscosity is plotted as a function of dimensional time for the infinite viscosity model given by Eq. (12) under quiescent conditions ($\dot{\gamma} = 0$). As shown in the figure, as $T_0$ decreases, viscosity increases at a faster rate. As a result, thixotropic behavior has been proposed to get weaker with increase in $T_0$.

We now discuss the limit of $T_0 \to \infty$ for both models. For the infinite viscosity model, in a limit of $t/T_0 \ll 1$, the Taylor series expansion of Eq. (12) leads to: $\eta_{inf}/\eta_0 = 1 + \alpha(t/T_0) + \sigma((t/T_0)^2)$, and $(\eta_{inf} - \eta_0)/\eta_0$ increases linearly with slope proportional to $1/T_0$. Accordingly, for the infinite viscosity model, it has been proposed that thixotropic behavior gets weaker with an increase in $T_0$. For finite viscosity model, in a limit of $t/T_0 \ll 1$, the Taylor series expansion of Eq. (11) leads to: $\eta_{fin}/\eta_0 = 1 + (\eta_p/\eta_0)(t/T_0) + \sigma((t/T_0)^2)$. Accordingly, for the finite viscosity model, $(\eta_{inf} - \eta_0)/\eta_0$ also increases linearly with slope proportional to $1/T_0$. In a limit of $T_0 \to \infty$ an increase in viscosity becomes so weak that it remains practically constant. Therefore, in this limit, viscosity in both the models: the infinite viscosity model as well as the finite viscosity model shows identical behavior (viscosity remains constant over the entire duration of the observation timescale). Paradoxically, this identical behavior in the limit of $T_0 \to \infty$, has been considered as having no (or weak) thixotropy limit by the infinite viscosity model while strong thixotropy limit by the finite viscosity model. The concept of thixotropic Deborah number also does not hold for this limit of $T_0 \to \infty$, as for any practically explorable observation times, $T_0/t \to \infty$, but thixotropic effect in this domain would actually be negligible. In addition, for thixo-viscoelastic material, the thixo-elastic parameter that is the ratio of stress relaxation time and $T_0$ will always be very small. However, there won't be dominance of thixotropy, and the material will always behave as a standard viscoelastic material.

The part of the definition of thixotropy suggests the phenomenon of an increase in viscosity is associated with the quiescent conditions. For the infinite viscosity model, as the name suggests viscosity diverges to infinity, while for the finite viscosity model viscosity always reaches the equilibrium value. Under the application of constant shear rate, applied to a system under a structureless state, viscosity does increase in both cases but with a reduced rate, and irrespective of the nature of the model, always reaches a steady state value. Therefore, how strongly viscosity increases with time can be considered as an indicator to judge the strength of the thixotropic character of a material under no-flow as well as constant shear rate conditions. Consequently, we obtain the derivative $d\ln\eta/d\ln t$ for both, the infinite viscosity model as well as the finite viscosity model. The corresponding expressions of $d\ln\eta/d\ln t$ for infinite viscosity model is given by:

$$\frac{d\ln\eta_{inf}}{d\ln t} = \alpha \frac{t}{T_0} exp\big(-\beta(T_0\dot{\gamma})(t/T_0)\big). \tag{13}$$

The limit of quiescent conditions $(T_0\dot{\gamma} \to 0)$ of Eq. (13) is given by

$$\frac{d\ln\eta_{inf}}{d\ln t} = \alpha \frac{t}{T_0}. \tag{14}$$

For the finite viscosity model $d\ln\eta/d\ln t$ is given by:

$$\frac{d\ln\eta_{fin}}{d\ln t} = \frac{\eta_p}{\eta_0}\frac{t}{T_0}\frac{e^{-\Lambda}}{\left[1 + \frac{\eta_p}{\eta_0}(1-e^{-\Lambda})\right]} exp\big(-\beta(T_0\dot{\gamma})(t/T_0)\big), \tag{15}$$

where $\Lambda$ is given by Eq. (5). Again in the no-flow limit $(T_0\dot{\gamma} \to 0)$, Eq. (15) reduces to:

$$\frac{d\ln\eta_{fin}}{d\ln t} = \left(\frac{t}{T_0}\right)\frac{\eta_p}{\eta_0}\frac{\exp\left(-\frac{t}{T_0}\right)}{\left(1 + \frac{\eta_p}{\eta_0}\left(1 - \exp\left(-\frac{t}{T_0}\right)\right)\right)}. \tag{16}$$

In Fig. 4, we plot $d\ln\eta/d\ln t$ as a function of $T_0/t$ for both the finite viscosity model (Eq. (16) with $\eta_p/\eta_0 = 100$) and the infinite viscosity model (Eq. (14) with $\alpha = 1$) for different $T_0\dot{\gamma}$ including the quiescent condition of $T_0\dot{\gamma} = 0$. The limit of $T_0/t \ll 1$ indicates that limit when the observation time is always significantly greater than $T_0$. On the contrary, $T_0/t \gg 1$ is that limit when observation time is extremely small compared to $T_0$. However, if we compare at any fixed observation time ($t$ = constant), change in $T_0/t$ indicates essentially the variation of $T_0$. It can be seen that, for the finite viscosity model, $d\ln\eta/d\ln t \to 0$ in both the limits $T_0/t \ll 1$ and $T_0/t \gg 1$ irrespective of the value of $T_0\dot{\gamma}$. The large $t$ or low $T_0$ branch as well as maximum associated with $d\ln\eta/d\ln t$ shift to higher value of $T_0/t$. More specifically under quiescent conditions, this behavior suggests that for the finite viscosity model, for any given time $t$, the thixotropic effect goes on diminishing in both high and low $T_0$ limit.

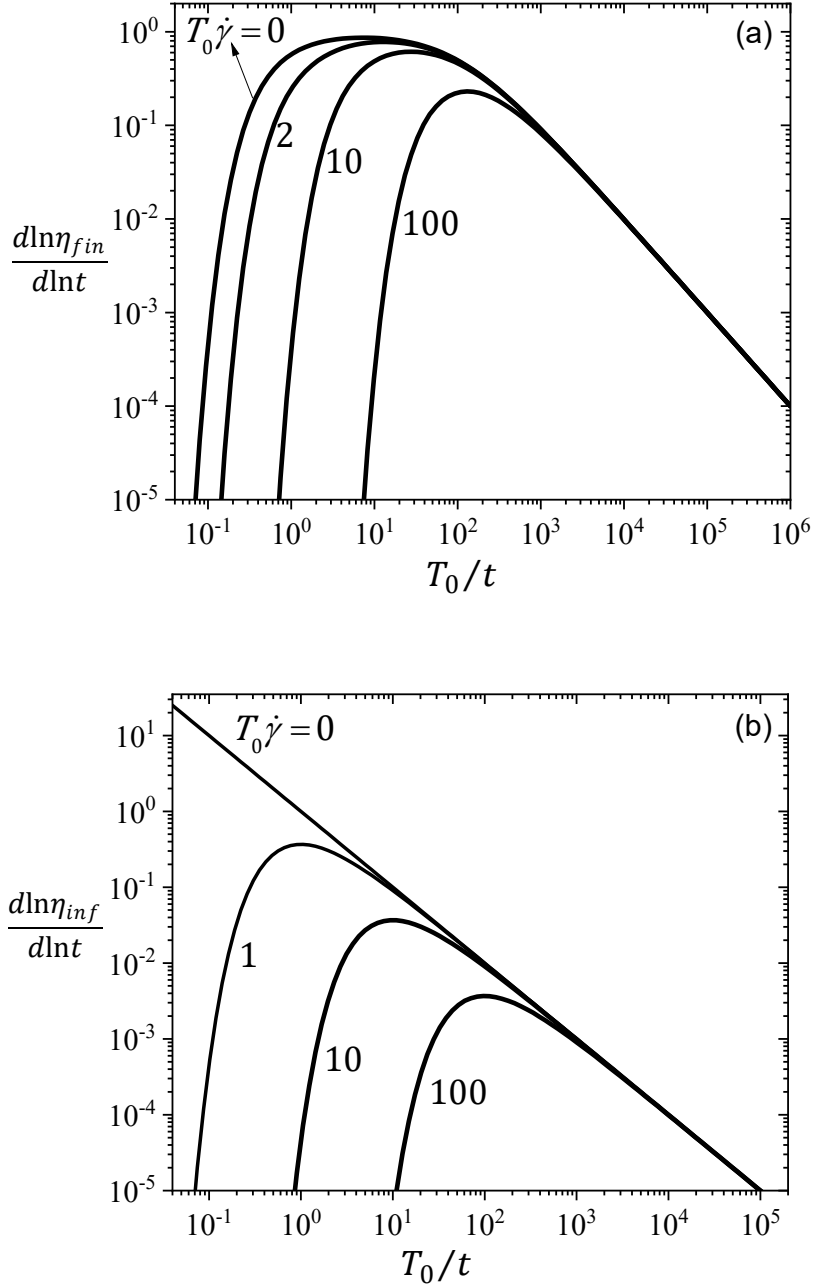

**Figure 4.** Rate of evolution of viscosity as a function of time ($d\ln\eta/d\ln t$) is plotted as a function of $T_0/t$ for different values of $T_0\dot{\gamma}$ for (a) the finite viscosity model (Eq. (16) with $\eta_p/\eta_0 = 100$) and (b) the infinite viscosity model (Eq. (14) with $\alpha = 1$).

On the other hand, for the infinite viscosity model for no flow conditions ($T_0\dot{\gamma} = 0$), $d\ln\eta/d\ln t \to \infty$ in a limit of $T_0/t \ll 1$ while $d\ln\eta/d\ln t \to 0$ in a limit of $T_0/t \gg 1$. This suggests that for the infinite viscosity model strength of the thixotropic phenomenon increases with decreasing $T_0$. For non-zero values of $T_0\dot{\gamma}$, however, the infinite viscosity model also attains a steady state. Consequently, in a limit of large times or $T_0/t \ll 1$, $d\ln\eta/d\ln t \to 0$ thereby showing a non-monotonic behavior. With increase in $T_0\dot{\gamma}$ the maximum in $d\ln\eta/d\ln t$ as well as large $t$ (or low $T_0$) branch shifts to lower values of $T_0/t$ such that as $T_0\dot{\gamma} \ll 1$, the maximum vanishes. Therefore, it can be seen that both models show qualitatively different behaviors under no-flow conditions.

It is important to note that the behavior shown in Fig. 2 is specifically for the models expressed by Eq. (14) and Eq. (16). Nevertheless, for no flow conditions, it is apparent that any model that incorporates finite viscosity will always show a non-monotonic behavior while any model that incorporates infinite viscosity, wherein viscosity continuously increases with time, will show a monotonically decreasing curve (may not be a straight line on a double logarithmic plot) when $d\ln\eta/d\ln t$ is plotted as a function of $T_0/t$ as shown in Fig. 4. This discussion, therefore, evidently suggests that limits associated with decreasing or increasing strength of thixotropy in terms of $T_0$ is applicable only when the equilibrium state associated with thixotropic material has infinite viscosity. Usually, such material does not attain equilibrium over practically explorable times, and hence viscosity keeps on increasing as a function of time indefinitely. More specifically, when viscosity keeps on increasing indefinitely, Fig. 4 suggests that an increase in $T_0$ weakens the thixotropic character of a material and the other way around. On the other hand, when material eventually achieves finite viscosity, no such correlation between change in $T_0$ and weakening/strengthening of the thixotropic character can be made. Therefore, specifically for the infinite viscosity structural kinetic model formalism given by Eq. (3) with $F(\Lambda) = \beta\Lambda$, Eq. (6), and Eq. (10), we can express two dimensionless numbers, $T_0/t$ and $T_0\dot{\gamma}$, as mentioned in table 1. For no flow condition, and when time $t$ is measured from the structureless state, greater is the dimensionless number $T_0/t$ weaker is the thixotropic effect at that time $t$. On the other hand, when system is subjected to shear rate $\dot{\gamma}$ in its structureless state, increase in $T_0\dot{\gamma}$ weakens the microstructural build-up and consequent viscosity evolution.

**Table 1.** Dimensionless numbers for the infinite viscosity structural kinetic model and associated limits

| $T_0/t$, No flow condition and time $t$ is measured from the structureless state | |
|---|---|
| $T_0/t \ll 1$ | Strong thixotropy limit |
| $T_0/t \gg 1$ | Weak or diminishing thixotropy limit |
| $T_0\dot{\gamma}$, when the system is subjected to a shear rate $\dot{\gamma}$ in its structureless state | |
| $T_0\dot{\gamma} \ll 1$ | Strong microstructural or viscosity evolution |
| $T_0\dot{\gamma} \gg 1$ | Weak or diminishing microstructural build-up or viscosity evolution |

**Thixotropic timescale from a generic expression of the structural kinetic model**

The methodology of defining the thixotropic timescale that was discussed in the previous section can be extended to a generic expression of the kinetic equation of the structural parameter $\bar{\lambda}$ bounded by $0$ (structureless state) and $1$ (fully structured equilibrium state) given by:

$$\frac{d\bar{\lambda}}{dt} = -k_1 f_{br}(\dot{\gamma}, \bar{\lambda}) g_{br}(t) + k_2 f_o(\dot{\gamma}, \bar{\lambda}) g_o(t) + k_3 f_p(\bar{\lambda}) g_p(t), \qquad (17)$$

where the pre-factors $k_1$, $k_2$, and $k_3$ are model parameters. The first term on the right is the flow-induced breakdown term, the second term is an orthokinetic term representing flow-induced collision leading to structural build-up and the third term is a perikinetic term that corresponds to Brownian/thermal motion-induced aggregation under no flow conditions [1, 7, 30, 31]. Correspondingly $f_{br}(\dot{\gamma}, \bar{\lambda}) g_{br}(t)$, $f_o(\dot{\gamma}, \bar{\lambda}) g_o(t)$ and $f_p(\bar{\lambda}) g_p(t)$ represent breakdown, orthokinetic and perikinetic functions. Most of the models use: $f_{br}(\dot{\gamma}, \bar{\lambda}) = \dot{\gamma}^a \bar{\lambda}^b$, $f_o(\dot{\gamma}, \bar{\lambda}) = \dot{\gamma}^c (1 - \bar{\lambda})^d$ and $f_p(\bar{\lambda}) = (1 - \bar{\lambda})^e$, where $a$, $b$, $c$, $d$ and $e$ are parameters. Furthermore, the independent time-dependent terms $g_{br}(t)$, $g_o(t)$ and $g_p(t)$ are not present in most of the models. However, since Dulart and Mewis [32] employ $g_{br}(t) = t^{-l}$, $g_o(t) = t^{-m}$ and $g_p(t) = t^{-n}$, with the model parameters $l = m = n$ in their formulation as it leads to stretched exponential relaxation behavior, we have included the same in a non-specific fashion in this generic expression. Moreover, strictly speaking, flow-induced structural build-up is not a thixotropic phenomenon, and hence for any further analysis we consider $k_2 = 0$. We now consider a special case of Eq. (17) with $g_{br}(t) = g_o(t) = g_p(t) = 1$, and carry out some rearrangement leading to:

$$\frac{1}{f_p(\bar{\lambda})} \frac{d\bar{\lambda}}{dt} = k_3 - k_1 \frac{f_{br}(\dot{\gamma}, \bar{\lambda})}{f_p(\bar{\lambda})}. \qquad (18)$$

As discussed before, since the structure parameter $\bar{\lambda} \in (0,1)$ represents a conceptual but vague notion of the extent of structure build-up, it can be easily converted to $\Lambda \in (0, \infty)$ for certain kinds of functional forms of $f_p(\bar{\lambda})$ that are usually used in the structural kinetic model by expressing:

$$d\Lambda = \frac{1}{f_p(\bar{\lambda})} d\bar{\lambda}, \qquad (19)$$

which can be used to rewrite Eq. (18) as:

$$\frac{d\Lambda}{dt} = k_3 - k_1 \frac{f_{br}(\dot{\gamma}, \bar{\lambda}(\Lambda))}{f_p(\bar{\lambda}(\Lambda))}. \qquad (20)$$

It can be seen that the form of Eq. (20) is equivalent to Eq. (3) (or Eq. (4), which is the same as Eq. (8)), that makes $T_0 = k_3^{-1}$ to be the thixotropic timescale.

Let us now consider another special case of Eq. (17) wherein $k_2 = 0$ and $g_{br}(t)$ and $g_p(t)$ are non-zero. Eq. (17) can be rearranged to give:

$$\frac{1}{f_p(\bar{\lambda})}\frac{d\bar{\lambda}}{dt} = k_3 g_p(t) - k_1 \frac{f_{br}(\dot{\gamma}, \bar{\lambda})}{f_p(\bar{\lambda})} g_{br}(t). \qquad (21)$$

In this equation, $\bar{\lambda} \in (0,1)$ can be converted to $\Lambda \in (0, \infty)$ by:

$$d\Lambda(t) = \frac{1}{f_p(\bar{\lambda})} d\bar{\lambda}(t) \qquad (22)$$

Incorporation of Eq. (22) into Eq. (17) leads to:

$$\frac{d\Lambda}{dt} = k_3 g_p(t) - k_1 \frac{f_{br}(\dot{\gamma}, \bar{\lambda}(\Lambda))}{f_p(\bar{\lambda}(\Lambda))} g_{br}(t). \qquad (23)$$

In this expression $k_3 g_p(t)$ has units of inverse of time. Consequently, it leads to a very interesting form of thixotropic timescale:

$$T_0(t) = \frac{1}{k_3 g_p(t)}, \qquad (24)$$

which suggests it depends on time and is not a constant as conventionally considered. This question, whether thixotropic timescale is necessarily a constant or should depend on observation time has not been considered in the literature and deserves a debate. The important fact is that the thixotropic behavior is inherently time-dependent, for which relaxation time has also been observed to be time-dependent [23, 33-37]. Hence, there should not be a constraint that prevents the thixotropic time from being time-dependent. Moreover, the perikinetic kinetic term in Eq. (17), which is responsible for thixotropic microstructural build-up, depends on the mobility of the constituents. With an increase in viscosity, the mobility of the same becomes sluggish and hence the thixotropic structural build-up also becomes slow. Therefore, consideration of a time-dependent thixotropic timescale could indeed be one of the ways to model a thixotropic system. If we consider $g_p(t) = t^{-n}$, $T_0(t) = k_3^{-1} t^n$. Such time dependent thixotropic timescale has been considered in the literature before and has indeed been observed to have power law dependence on time [24]. Under quiescent conditions, the structural kinetic model proposed by Dulart and Mewis [32] is given by: $\frac{d\bar{\lambda}}{dt} = k_3(1 - \bar{\lambda})t^{-n}$. Our analysis clearly shows that the thixotropic timescale for their model indeed depends on time and is given by: $T_0(t) = k_3^{-1} t^n$. By fitting their model to the experimental data, they obtain $n \approx 0.37$ and $n \approx 0.362$ respectively for their Fumed silica and Carbon black systems [32].

Physically thixotropic timescale suggests a timescale associated with microstructural evolution that causes an increase in viscosity of a material. However, microstructural evolution occurs over different length scales in a material and hence it is expected that it will occur over many timescales, leading to a spectrum of thixotropic times. The aforementioned proposal for the thixotropic timescale also paves the way for obtaining the spectrum of thixotropic timescales by generalizing the model proposed by Wei et al. [27, 38]. Accordingly, the structure parameter $\Lambda$ has been subdivided into an array of structure parameters $\Lambda_i$ having a relation:

$$h(\Lambda) = \sum_i w_i(t) h(\Lambda_i), \qquad (25)$$

where $h(\Lambda)$ is a suitable function of $\Lambda$ such as: $h(\Lambda) = \Lambda$ [27], $h(\Lambda) = 1 - e^{-\Lambda}$ [38], etc. In addition, each $\Lambda_i$ has an independent evolution expression given by:

$$\frac{d\Lambda_i}{dt} = \frac{1}{T_{0i}(t)} - c_i F(\Lambda_i, \dot{\gamma}), \qquad (26)$$

where $w_i(t)$ and $T_{0i}(t)$ represent the spectrum of thixotropic times. A pre-factor $c_i$, on the other hand, represents how individual $i^{th}$ structure parameter gets altered by the deformation field. While expressing the multimode model, we have gone one step further in expressing the evolution of $\Lambda_i$ by considering the time-dependent spectrum of thixotropic times. The expressions equivalent to Eqs. (25) and (26), less the deformation field induced rejuvenation term, have been routinely used in the glassy polymer literature [39-41], wherein the physical aging part of the same is equivalent to thixotropic structural build-up. The corresponding KAHR model (Kovacs-Aklonis-Hutchinson-Ramos model) also considers $\sum_i w_i = 1$ [39, 41]. The thermo-rheological simplicity is maintained by keeping $w_i$ to be a constant [39-41]. If a material behavior does not require the spectrum of thixotropic timescales to depend on time, $w_i$ and $T_{0i}$ can be considered as constants. To complete the model Eqs. (25) and (26) can be combined with a constitutive equation (such as the generalized Newtonian Model given by Eq. (6)) and an expression that relates parameters of the constitutive equation and $\Lambda$ (such as given by Eq. (9) or (10)). The parameters $w_i(t)$, $T_{0i}(t)$ and $c_i$ including their time dependence, if any, can be obtained by fitting the model to the experimental data. The overall framework can be extended to include viscoelasticity in the thixotropic framework by using an appropriate linear or non-linear viscoelastic model with a single mode or spectrum of relaxation times. It would be interesting to understand whether there could be a relation between $\Lambda_i$ (and $T_{0i}$th mode) and the viscoelastic relaxation mode. However, in case the viscoelasticity is included, the rejuvenation term in Eq. (26) needs to be explicitly dependent only on stress and not on the deformation rate in order to prevent violation of the second law of thermodynamics [29]. However, in the present work, we restrict ourselves to inelastic constitutive equations.

**The Equilibrium state of a thixotropic material: Finite Viscosity or Infinite Viscosity:**

The above discussion raises an important question on how under quiescent conditions the real thixotropic materials behave in a limit of long times and the nature of the equilibrium state associated with the same. There are varied opinions in the literature on this matter. Moller et al. [42] term thixotropy to be a phenomenon associated with reversible physical aging and shear rejuvenation, such that under rest material develops a structure that is flow-resisting (or shows yield stress). They also state yield stress of thixotropic material necessarily increases with time. This suggests that Moller et al. [42] consider thixotropy necessarily leads to infinite viscosity eventually. Balmforth et al. [43] also take a similar stand wherein they mention that thixotropy implies an increase in static yield stress as a function of rest time. Others consider thixotropic behavior to be more accommodating as many employ different kinds of finite viscosity models in their formulations [5, 7, 8] indicating they are comfortable with thixotropic materials having an equilibrium state that has finite viscosity. The topic of thixotropy and its relation to the presence of yield stress (suggesting that viscosity indeed diverges) has been dealt with within the literature in great detail and varied opinions exist on this matter that have been summarized elsewhere [4, 33, 44, 45].

In the literature, a huge number of systems have been termed as thixotropic, and an exhaustive list of the same is given by Barnes [11], Mewis and Wagner [1, 22, 46], and Larson and Wei [7]. To the best of our knowledge and understanding, these systems can be divided into two categories. One category for which there is enough evidence in the literature that under quiescent conditions the viscosity of the same goes on increasing indefinitely. In the case of the other category, the experiments have not been performed to ascertain whether the material reaches equilibrium and viscosity reaches a time-invariant constant value that is finite. Many times, a potential thixotropic material shows a constant value of $G'$ and $G''$ as a function of time when subjected to small amplitude oscillatory shear [4]. However, in our opinion, that does not necessarily mean the material is time-invariant. We believe that the most reliable way to ascertain whether rheological behavior is time-dependent or not is to subject the same to either step stress (leading to creep flow) or step strain (inducing stress relaxation) at different intervals of time $(t_w)$ [3, 47]. In this case, since $t$ is time, $t - t_w$ becomes the time elapsed since application of strep stress or step strain. If the corresponding creep compliance $(J)$ or stress relaxation modulus $(G)$ shows an additional dependence on $t_w$, such that $J = J(t - t_w, t_w)$ or $G = G(t - t_w, t_w)$, it clearly indicates that material does not reach an equilibrium over a period of $t_w$ [33]. We believe that unless this test has been performed, it would not be appropriate to say that any material has reached time invariant equilibrium state.

Recently Joshi and coworkers [4, 48, 49] showed that while aqueous Carbopol dispersion and hard-sphere glass (59.5 and 62.5 volume % suspension of sterically stabilized spherical poly(methyl methacrylate) in squalene) do not show any evolution of dynamic moduli as a function of time, these systems show a strong waiting time dependence in creep: $J = J(t - t_w, t_w)$, and hence have been termed to be thixotropic. Furthermore, many materials that have been termed thixotropic in the literature [1, 7, 11, 22, 42, 46] show yield stress, which itself is a manifestation that their viscosity diverges with an increase in time. The corresponding value of yield stress may remain constant or increase with time. In our opinion, the constant value of yield stress, however, does not imply that the material has reached the equilibrium state as it can still show waiting time-dependent creep compliance $(J = J(t - t_w, t_w))$, as observed by Bhattacharyya et al. [4].

The above discussion brings us to an important question whether there exists a thixotropic material that shows an equilibrium state and whether such an equilibrium state has finite viscosity. Let us consider a thought system that constitutes a dilute suspension of colloidal particles that undergo Brownian motion, which share attractive interactions. With time, the particles are expected to form clusters, whose size grows with time. Such microstructural build-up is a characteristic thixotropic feature as it indeed leads to an increase in viscosity as a function of time under no flow/no stress conditions. It has been claimed that for sufficiently strong attractive interaction energy, gel formation that spans the space – that is suggestive of divergence of viscosity – may happen for arbitrary low-volume fractions ($\phi \to 0$) [50]. The literature, however, is mute, if such percolated gel formation should take place even if the particles possess just enough attractive interactions capable of forming clusters. We, therefore, believe that theoretically there could a very small minority of thixotropic systems that show an equilibrium state, whose viscosity is finite, and hence can be modeled by the finite viscosity model expressed by Eq. (9). Since such materials will always be inelastic, increase and decrease in their viscosity respectively under quiescent and under deformation field conditions will lead to their recognition as thixotropic. Nevertheless, in such systems, change in viscosity with time under quiescent conditions as well as under the application of the deformation field would be extremely small, and hence the change in viscosity may not be easily detectable. We further believe that most of the systems of practical importance undergo continuous evolution of viscosity with time over reasonably explorable timescales such that their viscosity eventually diverges, and the equilibrium state if it is realizable, has infinite viscosity. Consequently, the material necessarily shows yield stress through viscosity bifurcation [4, 16]. For such systems, the behavior of $J(t - t_w, t_w)$ or $G(t - t_w, t_w)$ shows classical signature for thixotropic materials that is different from non-thixotropic viscoelastic materials. Agarwal et al. [3] leverage this difference to propose a criterion that distinguishes a viscoelastic response from a thixotropic response.

In the previous section, we discussed the definition of thixotropic timescale from the point of view of generic structural kinetic model formalism. It is clear that the definition is not affected by whether the equilibrium state of a material possesses finite viscosity or infinite viscosity. However, the viscosity of practically important thixotropic materials seems to diverge with time. For such materials, we can certainly mention that an increase in the thixotropic timescale does make the thixotropic effect weaker. If the thixotropic timescale increases with time as indicated by Eq. (24), the thixotropic behavior of a material gets weaker as time passes.

**A phenomenological proposal for thixotropic timescale**

The earlier section discussed a universal expression for the thixotropic timescale, along with its corresponding distribution, for a generic kinetic expression of the evolution of structure parameter. However, two main concerns persist. Firstly, determining a thixotropic timescale becomes challenging when a material's experimental behavior doesn't follow the predictions of a structural kinetic model outlined in Eq. (17) for single mode or Eq. (26) for multiple modes (The way multimode Maxwell model is guaranteed to predict any linear viscoelastic response, which leads to the determination of relaxation time distribution, it is not clear whether the multimode structural kinetic model is equipped to predict any thixotropic response). Secondly, whether calculated by Eq. (17) or (26), the thixotropic timescale still relies on a structural parameter that lacks clear physical significance until it is linked to rheological parameters. Consequently, the same spectrum will lead to different rheological responses if different relationships between rheological parameters and structure parameters are used. These aspects indeed raise doubts about the reliability of the estimated thixotropic timescale and its distribution. Unfortunately, there is no definitive resolution available for these concerns at the moment particularly when adhering to a structural kinetic model to determine the thixotropic timescale. We believe the other way to define the thixotropic timescale is by taking a phenomenological approach. Considering the fundamental definition of thixotropy, which involves an increase in viscosity under quiescent conditions, the thixotropic timescale can be defined by expressing the build-up part of the viscosity (under no flow conditions) as:

$$\frac{1}{\eta}\frac{d\eta}{dt} = \frac{1}{\tau_{thix}} \quad \text{for } \dot{\gamma} = 0. \tag{27}$$

Therefore, if $\eta_{NF} = \eta|_{\dot{\gamma}=0}$ is viscosity under no-flow conditions, Eq. (27) leads to an explicit definition of the thixotropic timescale given by:

$$\tau_{thix} = \left(\frac{d\ln\eta_{NF}}{dt}\right)^{-1}, \tag{27a}$$

which is intrinsic to the evolution of viscosity under quiescent conditions – the fundamental basis of thixotropic behavior. There are several important aspects associated with the proposed expression of thixotropic timescale. Firstly, the phenomenological basis of the same wherein the expression allows estimation of the thixotropic timescale directly from the experimental data. Consequently, the proposed expression is independent of any structural kinetic formalism accompanied by a constitutive equation or other mathematical model used to describe the thixotropic behavior. The next aspect is that the proposed definition given by eq. (27) does not assume the thixotropic timescale to be a constant, but allows its variation as a function of time depending upon the evolution behavior of viscosity. As discussed before, thixotropy, by definition, is a time-dependent phenomenon, wherein the structure builds up with time. Essentially, in many cases, it is the length scale associated with the structure that progressively increases with time. An increase in length scale is expected to decrease the mobility of the structure, thereby increasing the timescale associated with the thixotropic phenomenon.

We believe that experimental estimation of the proposed thixotropic timescale by Eq. (27) that requires the determination of viscosity under no-flow conditions is, in principle, a straightforward task. If thixotropic material is inelastic, the evolution of viscosity can be monitored by assessing the evolution of viscous modulus, as in the terminal regime $\eta \approx G''/\omega$. Furthermore, as a material undergoes structural build-up with time, it gains viscoelastic character, and it is likely that it eventually shows yield stress. We believe that the information on viscosity enhancement under such circumstances can be obtained by monitoring the evolution of instantaneous modulus ($G_0 = G(t \to 0)$) and relaxation time of the same ($\tau_{sr}$, where subscript $sr$ stands for stress relaxation) as viscosity is a product of instantaneous modulus and relaxation time: $\eta \approx G_0 \tau_{sr}$. Both modulus and relaxation time can be obtained by subjecting the thixotropic material to step strain or step stress at different waiting times. This procedure to estimate modulus and relaxation time has been discussed in detail elsewhere [3, 4, 34, 51]. Usually, the stress relaxation time of the materials that have been considered thixotropic such as clay dispersion, hard-sphere glass, concentrated emulsions, etc. is observed to show power law dependence on time given by: $\tau_{sr} = A\tau_m^{1-\mu} t^\mu$, where $A$ is a proportionality constant, $\tau_m$ is the microscopic timescale and $\mu$ is the power law index [3, 4, 34, 51]. In this expression $t$ is the time associated with evolution of material from its structureless state. Usually, the instantaneous modulus is a weak function of time, and for the sake of simplicity, if we consider it to be a constant, we get: $\eta \approx G_0 A \tau_m^{1-\mu} t^\mu$. The corresponding thixotropic timescale can then be estimated as: $\tau_{thix} \approx t/\mu$. It is interesting to note that the phenomenological way of estimating thixotropic timescale shows a linear dependence on time.

In addition to experimental determination, the proposed phenomenological expression of the thixotropic timescale can be used for any theoretical formalisms. The formalisms that employ non-structural kinetic model approaches such as the population balance approach [52], the estimation of the derivative given by Eq. (27) will directly lead to the thixotropic timescale. On the other hand, for the structural kinetic models, the derivative given by Eq. (27) can be expressed in terms of the parameters of the same. For example for the finite viscosity model given by Eq. (9), the phenomenological thixotropic timescale can be obtained by Eq. (16) by simply multiplying both the sides by $t$ and taking the inverse. It clearly shows that thus defined phenomenological thixotropic timescale depends on time and is equal to $\tau_{thix} = T_0 \eta_0 / \eta_p$ in a limit $t/T_0 \ll 1$ while $\tau_{thix} \to \infty$ in a limit $t/T_0 \gg 1$. or For the infinite viscosity model given by Eq. (10) the phenomenological thixotropic timescale is given by Eq. (14), which interestingly is a constant given by $\tau_{thix} = T_0/\alpha$. As mentioned before, the infinite viscosity model given by Eq. (10) necessarily shows yield stress through viscosity bifurcation.

The proposed phenomenological definition of thixotropic timescale given by Eq. (27) suggests that smaller the value $\left(\frac{d\ln\eta_{NS}}{dt}\right)^{-1}$ is, stronger is the thixotropic character of a material at that time and vice a versa. The phenomenological definition of thixotropic timescale therefore leads to a natural dimensionless number and is expressed as a ratio of $\tau_{thix}$ to $t$ given by:

$$\frac{\tau_{thix}}{t} = \frac{1}{t}\left(\frac{d\ln\eta_{NF}}{dt}\right)^{-1} = \left(\frac{d\ln\eta_{NF}}{d\ln t}\right)^{-1}, \qquad (28)$$

where $t$ is the time associated with the evolution of material since its structureless state. If material is shear melted, $t$ is the time elapsed since the cessation of shear melting. This ratio $\tau_{thix}/t$ signifies the importance of thixotropic viscosity build-up at any given time $t$. Smaller the value of $\tau_{thix}/t$ or $\left(\frac{d\ln\eta_{NS}}{d\ln t}\right)^{-1}$ is, stronger is the thixotropic character of material at that time $t$. Consequently, for $\left(\frac{d\ln\eta_{NS}}{d\ln t}\right)^{-1} \gg 1$ effect of thixotropy can be neglected at that time $t$. While our proposed definition of the thixotropic timescale is associated with the no-flow condition, under the application of flow field viscosity may undergo an increase at a reduced rate compared to that under quiescent conditions. Representative behavior of $\frac{d\ln\eta}{d\ln t}$ under the application of constant $\dot{\gamma}$, shown in Fig. 4, essentially suggests this behavior. However, if material in a structured initial state (high viscosity) is subjected to a strong flow field, the viscosity is expected to decrease with time and $\left(\frac{d\ln\eta}{dt}\right)^{-1}$ will have a negative value. This will correspond to a breakdown timescale of a material that will depend on the initial state of a material and the nature and magnitude of the applied deformation field.

**Conclusions**

The thixotropic timescale represents the timescale associated with the thixotropic phenomenon, which essentially involves microstructural build-up that leads to an increase in viscosity. In the literature, various ways have been prescribed to obtain the thixotropic timescale, the most prominent of which is its estimation from a specific form of kinetic expression for the evolution of structural parameter. In this work, we begin by analyzing two competing expressions for the evolution kinetics of structure parameters that seemingly lead to different limits of thixotropic timescales. In one case, a structure parameter is constrained to the range 0 (structureless state) and 1 (fully structured equilibrium state), whereas in the other case, it spans 0 (structureless state) to ∞ (fully structured equilibrium state). Very interestingly, careful assessment of both kinetic expressions leads to the identical meaning of the thixotropic timescale. Considering one of the crucial aspects of the structural kinetic model, which is a relationship between structure parameter and rheological parameters, particularly the viscosity, we consider two models, the finite viscosity model and the infinite viscosity model. The former, as the name suggests, assumes the viscosity of the equilibrium state to be finite. The latter, on the other hand, prescribes the viscosity of the same to be infinite in a limit of the equilibrium state. A meticulous assessment of both the model formulations suggests that we can relate a variation in the thixotropic timescale to a change in the strength of the thixotropic character of a material only when the viscosity of the same continuously increases with time and eventually diverges under quiescent conditions. Particularly, when viscosity diverges over time in the absence of flow, which we believe is the most practical scenario, our study indicates that the thixotropic phenomenon is attenuated with an increase in the thixotropic timescale.

Subsequently, we consider the most generic expression for the kinetics of the structure parameter evolution, and by performing appropriate manipulation of the same, propose a universal candidate for the thixotropic timescale that is invariant of the nature of the evolution equation. Interestingly, the proposed methodology can be easily extended to encompass a spectrum of thixotropic timescales. Moreover, the methodology can also adapt to incorporate the prospect of time dependency in thixotropic timescales, a proposal that needs to be debated in the literature. Given that thixotropy is a time-dependent phenomenon, where structure builds up over progressively larger lengthscales, it appears reasonable to consider the plausibility of time-dependent thixotropic timescales. Although we establish the thixotropic timescale(s) using a generic equation describing the kinetics of structural parameter evolution, the ambiguity persists because the structural parameter provides only an abstract idea of the material's state, necessitating an explicit quantification through its relationship with viscosity. To tackle this issue, we introduce a new phenomenological version of the thixotropic timescale given by, $\tau_{thix} = (d\ln\eta/dt)^{-1}$, where $\eta$ is viscosity and $t$ is time, which relies on time-dependent changes in viscosity. Interestingly, this approach is universally applicable, catering to both experimental and theoretical frameworks. The introduction of this definition not only streamlines the process of identifying the thixotropic timescale through experimental/theoretical investigations but also maintains coherence with the conventional notion of thixotropy, wherein the gradual increase in viscosity as a function of time in the absence of flow holds significance.

**Acknowledgment:** The author acknowledges financial support from the Science and Engineering Research Board, Government of India (Grant numbers: CRG/2022/004868 and JCB/2022/000040). He thanks Professor Sachin Shanbhag for constructive comments. He is also indebted to Professor Ronald Larson for critical feedback and discussion on various aspects of thixotropy and its timescale.